\documentclass[twocolumn,prl,tightenlines,superscriptaddress,showpacs]{revtex4-1}

\usepackage{amsmath}
\usepackage{amssymb,amsfonts,latexsym}
\usepackage{bm}
\usepackage[mathcal]{euscript}
\usepackage{graphicx}
\usepackage{epsfig}
\usepackage{color}
\usepackage{xfrac}
\usepackage{mathrsfs}

\usepackage{hyperref}
\usepackage{url}

\begin{document}

\title{Breakdown of Ergodicity and Self-Averaging in Polar Flocks with Quenched Disorder}

\author{Yu Duan}
\affiliation{National Laboratory of Solid State Microstructures and Department of Physics, Collaborative Innovation Center of Advanced Microstructures, Nanjing University, Nanjing 210093, China}

\author{Beno\^{\i}t Mahault}
\affiliation{Max Planck Institute for Dynamics and Self-Organization (MPIDS), 37077 G\"ottingen, Germany}

\author{Yu-qiang Ma}
\affiliation{National Laboratory of Solid State Microstructures and Department of Physics, Collaborative Innovation Center of Advanced Microstructures, Nanjing University, Nanjing 210093, China}

\author{Xia-qing Shi}
\affiliation{Center for Soft Condensed Matter Physics and Interdisciplinary Research, Soochow University, Suzhou 215006, China}

\author{Hugues Chat\'{e}}
\affiliation{Service de Physique de l'Etat Condens\'e, CEA, CNRS Universit\'e Paris-Saclay, CEA-Saclay, 91191 Gif-sur-Yvette, France}
\affiliation{Computational Science Research Center, Beijing 100193, China}

\date{\today}

\begin{abstract}
We show that spatial quenched disorder affects polar active matter in ways 
more complex and far-reaching than believed heretofore. 
Using simulations of the 2D Vicsek model subjected to random couplings or 
a disordered scattering field, 
we find in particular that ergodicity is lost in the ordered phase, the nature of which we show to depend
qualitatively on the type of quenched disorder: for random couplings,
it remains long-range ordered, but qualitatively different from the pure (disorderless) case. 
For random scatterers, polar order varies with system size but
we find strong non-self-averaging, with
sample-to-sample fluctuations dominating asymptotically, which prevents us from elucidating the asymptotic
status of order.
\end{abstract}

\maketitle

Spatial quenched disorder is known to be able to affect qualitatively the asymptotic properties of various systems
\cite{imry1975random-field,edwards1975theory,young1998spin,vink2006critical,aizenman1989rounding,harris1974effect,grinstein1976application, aharony1984multicritical,fisher1997stability,tissier2006unified}.
Its influence on active matter has recently attracted interest,
and rightly so since in many of the corresponding real situations active particles have to avoid obstacles,
or move on a rough substrate or in a disordered mesh \cite{bechinger2016active}. 

While some interesting results were obtained for scalar active matter 
\cite{reichhardt2014active,pince2016disorder-mediated,sandor2017dynamic,morin2017diffusion,reichhardt2018clogging,dor2019ramifications,ro2020disorder},
many of these studies have dealt with the case of dry polar flocks,
in continuity with the seminal
role played by the Vicsek model and the Toner-Tu theory \cite{vicsek1995novel,toner1995long,toner1998flocks,toner2012reanalysis,marchetti2013hydrodynamics,chate2020dry}.  
Most efforts were devoted to the fate of the two-dimensional (2D)
ordered liquid phase moving on some random substrate. 
It was found that an optimal amount of noise or disorder can maximize polar order \cite{chepizhko2013optimal,chepizhko2015active,quint2015topologically,martinez2018collective}.
Experiments studied how flocks of Quincke rollers found in 
\cite{bricard2013emergence} are altered and 
eventually destroyed by quenched disorder \cite{morin2016distortion,chardac2020meandering}.
Recently, Toner and Tu \cite{toner2018swarming,toner2018hydrodynamic} extended their theory of the homogeneous ordered phase to take quenched disorder into account, 
predicting in particular quasi-long-range order in 2D. 
Numerical work has produced partial results compatible with these predictions \cite{chepizhko2013optimal,chepizhko2015active,toner2018hydrodynamic,das2018polar,das2018ordering}. 

The study of disordered systems has a long history outside
active matter. 
Important concepts in this context are ergodicity and self-averaging, which can both be broken by disorder.
Ergodicity is lost when multiple configurations coexist for a given sample (realization of disorder). 
Systems for which spatial and sample averages are not equivalent in the thermodynamic limit are non-self-averaging
\cite{wiseman1995lack,aharony1996absence,vojta2006rare,fish2010nematics,pigeard2011loss,russian2017self-averaging,corberi2020quasideterministic}. 
It is also known that the type of quenched disorder can make a difference
\footnote{
A classic example is the hyperscaling relation at play at criticality: it is clearly violated in the random-field Ising model,
but remains valid for the random-bond Ising model \cite{fisher1974the,berche2004bond,vink2010finite-size}.}.
Somewhat surprisingly, ergodicity, self-averaging, and the influence of the type of quenched disorder have all been largely ignored in the active matter studies published so far
\footnote{Exceptions are found in \cite{peruani2018cold}, 
where the effects of the type of disorder were considered but mostly on noiseless, non- or weakly-interacting active particles deep in the disordered phase,
and in the recent preprint \cite{chardac2020meandering}, which we discuss at the end of this work.}.

In this Letter, we 
show that quenched disorder affects polar active matter in ways 
more complex and far-reaching than believed heretofore. 
Using simulations of the 2D Vicsek model, we find that
quenched disorder breaks ergodicity and rotational invariance in the ordered phase: several dynamical attractors
coexist for a given realization of disorder.
In the disordered phase, ergodicity is recovered, but the short correlation length dynamics are organized around an underlying sample-dependent skeleton best revealed in time-averaged fields. 
The type of disorder applied does not influence the above properties,  
but it can fundamentally change the structure of the phase diagram, self-averaging, 
and the nature of the ordered phase:
A random coupling- (or noise-) strength landscape does not alter the phase diagram and 
yields a self-averaging long-range ordered phase, albeit different from the Toner-Tu liquid of the pure case. 
Random scatterers, on the other hand, deeply modify the layout of the phase diagram,
and leaves non-ergodic and non self-averaging ordered regimes where
3 types of fluctuations compete 
(dynamical/thermal, sample-to-sample, but also between attractors existing for a given sample), 
a numerically challenging situation that prevents
us from elucidating the asymptotic nature of this phase.

%%%%%%%%% definition of micro models %%%%%%%%%%%%
We consider extensions of the standard Vicsek model (VM) with angular noise \cite{vicsek1995novel,chate2008collective,solon2015phase}. 
Like in the VM,
particles $i=1,...,N$ with position ${\bf r}_i$ and orientation ${\bf e}_i$ 
move at discrete timesteps with constant speed $v_0$:
${\bf r}_i^{t+1} = {\bf r}_i^t + v_0 {\bf e}_i^{t+1}$.
They locally align their velocities with neighbors, but they evolve on a static disordered landscape 
that influences their motion. Here we present results obtained with two types of such quenched spatial disorder
hereafter called random couplings (RC) and random scatterers (RS).
We use square domains of linear size $L$ with periodic boundary conditions, divided into unit boxes 
in which quenched disorder variables are defined. Orientations are governed by one of the following equations:
\begin{subequations}
\begin{align}
\label{eq:VM-RT}
{\bf e}_i^{t+1} &=\left( {\cal R}_\varepsilon^n \circ {\cal R}_\eta^{i,t} \circ  {\cal U} \right) \left[ \langle {\bf e}_j^{t} \rangle_{j\sim i} \right] \;\; & {\rm (RS)} \\
\label{eq:VM-RP}
{\bf e}_i^{t+1} &=( {\cal R}_{\eta(n)}^{i,t} \circ  {\cal U} ) \left[ \langle {\bf e}_j^{t} \rangle_{j\sim i} \right] \;\; & {\rm (RC)}
\end{align}
\label{eq:micro_model}
\end{subequations}
where $n$ is the index of the unit box containing ${\bf r}_i^t$, 
$\langle . \rangle_{j\sim i}$ is the average over all particles $j$ within unit distance of $i$ (including $i$), and
${\cal U}[{\bf u}] = {\bf u}/|{\bf u}|$ returns unit vectors. 
In the RS case, ${\cal R}_\eta^{i,t}[{\bf u}]$ is the angular noise of the standard VM ---it rotates vector {\bf u} by a random angle drawn for each particle $i$ at each timestep $t$ from a uniform distribution inside 
an arc of length $2\pi\eta$ centered on ${\bf u}$---, 
while ${\cal R}_\varepsilon^n$ rotates vectors by some fixed angle defined initially on each box $n$,
drawn from a zero-mean uniform distribution of width $2\pi\varepsilon$. 
In the RC case, ${\cal R}_{\eta(n)}^{i,t}$ is similar to ${\cal R}_\eta^{i,t}$, 
but the noise amplitude $\eta(n)$, different in each box $n$, 
is drawn once and for all from a uniform distribution over $[\eta,\eta+\varepsilon]$.
Note that the above models reduce to the ``pure", disorderless VM for $\varepsilon=0$.

{\it Finite-size phase diagrams.} 
The VM is well known to be governed by two main parameters, 
the global number density of particles $\rho_0$ 
and the (annealed) noise strength $\eta$. 
For large $\rho_0$ or small $\eta$, a polar liquid with true long-range order is observed, 
whereas only a disordered gas (with short-range correlations) exists at low densities and strong noise.
In the $(\rho_0,\eta)$ plane, these two phases are separated by a coexistence domain in which
dense, ordered bands travel in a sparse gas \cite{gregoire2004onset,solon2015phase,chate2020dry}.

%%%%%%%%%%%%%%%%%%%%%%%%%%%
\begin{figure}
	\includegraphics[width=\columnwidth]{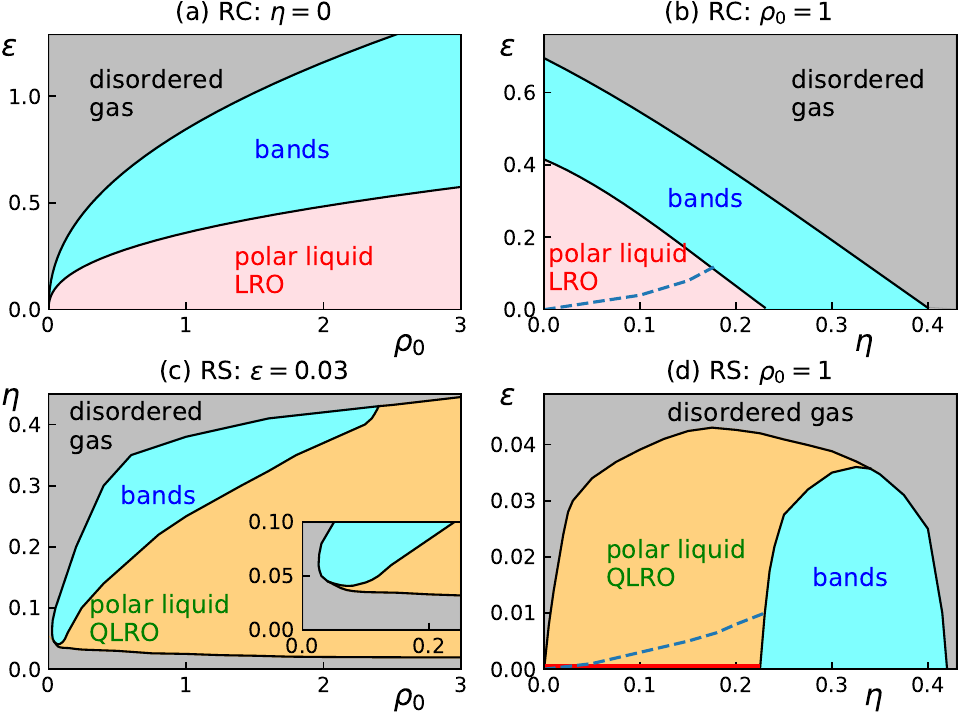}
	\caption{Stylized finite-size phase diagrams drawn from the data presented in \cite{Note3}.
	Top: RC case: (a) $(\rho_0,\varepsilon)$ plane for $\eta=0$;
	(b) $(\eta,\varepsilon)$ for $\rho_0=1$.
	Bottom: RS case: (c) $(\rho_0,\eta)$ plane at fixed $\varepsilon=0.03$ 
	(the inset shows the small-$\rho_0$, small-$\eta$ region);
	(d) $(\eta,\varepsilon)$ for $\rho_0=1$.
	The blue dashed line in (b) and (d) marks ergodicity-breaking at the system size considered.
	The pure VM ($\varepsilon=0$) lies on the x-axis in (a,b,d).
	 }
	\label{FIG1}
\end{figure}
%%%%%%%%%%%%%%%%%%%%%%%%%%%

A detailed study of the 3-parameter $(\rho_0,\eta,\varepsilon)$ phase diagrams of our RS and RC quenched disorder models 
is a very demanding task. 
Our efforts have led to the `stylized' finite-size phase diagrams presented in Fig.~\ref{FIG1} 
(the protocol followed to define them is detailed in \footnote{See Supplementary Information at ...}). 
The RC phase diagram in the $(\rho_0,\varepsilon)$ plane, but also in the $(\rho_0,\eta)$ plane (not shown),
looks identical to that of the VM (Fig.~\ref{FIG1}(a,b)).  
On the other hand,  
quenched disorder substantially modifies the layout in the RS case (Fig.~\ref{FIG1}(c,d)): 
the bands region remains present but its extent 
is bounded away from both low and high $\rho_0$ or $\eta$ values at any finite $\varepsilon$.
The ordered region is also bounded similarly. 
These RS results extend and clarify the findings of \cite{chepizhko2013optimal,chepizhko2015active}.

We now turn to the characterization of the encountered phases, focussing on
ergodicity, self-averaging, fluctuations, and memory. 
We use the modulus and direction of the instantaneous global polar order,   
$m=|\langle {\bf e}_j^t\rangle_{j}|$ and $\theta_m=\arg\langle {\bf e}_j^t\rangle_{j}$,
as well as coarse-grained fields calculated on the unit boxes on which quenched disorder is defined, notably the
momentum field ${\bf m}({\bf r},t)$. All numerical details can be found in \cite{Note3}.
All results presented below were obtained for $\rho_0=1$, as in Fig.~\ref{FIG1}(b,d).

Most phases are qualitatively different from those of the disorderless case.
The only exception is the coexistence phase: with any type of disorder, we found that its characteristic 
traveling bands retain their main properties, and notably lead to global long-range order \cite{TBP}.

%%%%%%%%%%%%%%%%%%%%%%%%%%%
\begin{figure}[t!]
	\includegraphics[width=\columnwidth]{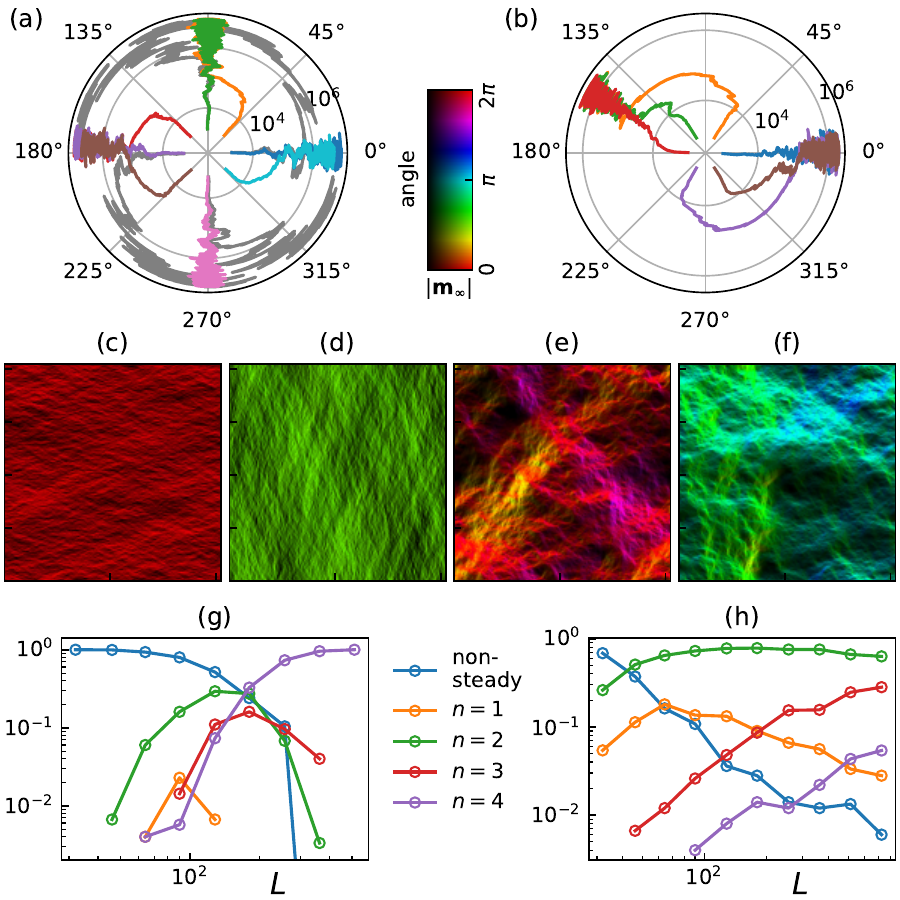}
	\caption{Ergodicity breaking in the RC (left, $\eta=0$, $\varepsilon=0.2$) and 
	RS (right, $\eta=0.18$, $\varepsilon=0.035$) ordered phases ($L=2048$ in (a-f)).
	(a,b) $\theta_m(t)$ observed on a single sample from initial conditions ordered 
	along 8 (left) and 6 (right) different directions. 
	Time increases radially outward (log scale) for 2 million timesteps. Grey curves: pure case (VM, $\varepsilon=0$).
	(c,d) long-time average momentum field ${\bf m}_\infty({\bf r})$ of 2 of the RC-case attractors shown in (a) (colormap in top row).
	(e,f) same as (c,d) but for RS-case attractors of (b).
	(g,h) fraction of non-steady and 1-, 2-, 3-, 4-attractor samples vs system size.
	 }
	\label{FIG2}
\end{figure}
%%%%%%%%%%%%%%%%%%%%%%%%%%%

{\it Ergodicity} is broken by quenched disorder in the globally ordered regimes found at finite size: 
for a typical realization of disorder (sample) of a large-enough system,
different initial conditions typically lead to different polarly ordered steady states. 
However, one finds only a rather small number of these attractors which
each attract many different initial conditions (Fig.~\ref{FIG2}). 
Each attractor is best characterized by the long-time-average of momentum field, 
${\bf m}_\infty({\bf r}) = \lim_{T\to\infty} {\bf m}_T({\bf r})$ with 
${\bf m}_T({\bf r}) = \tfrac{1}{T} \sum_{t=t_0}^{t=t_0+T} {\bf m}({\bf r},t)$.
However $\theta_m$
remains quasi-constant in time in most cases, and different from attractor to attractor, so that following 
it is sufficient to distinguish them. 
Quenched disorder thus fixes global order at particular angles, in contrast with the pure, disorderless case, for which
$\theta_m$ wanders slowly in a diffusive manner (Fig.~\ref{FIG2}(a,b)).

To be true, global order continues to wander in small systems with weak quenched disorder,
presumably because then no local configuration of disorder is strong enough to pin $\theta_m$.
For a given sample, there exists, within the ordered phase, 
an $L$-dependent region bordering the $\varepsilon=0$ axis inside which ergodicity is not yet broken,
located below the dashed lines in Fig.~\ref{FIG1}(b,d). 
Increasing $L$, this region shrinks:
 `non-steady', i.e. ergodic, samples dominate
at small size, but their fraction quickly decreases, 
while more and more attractors are found on average (Fig.~\ref{FIG2}(g,h)). 

While the above observations hold for both RC and RS disorder, there are important differences, notably
in the spatial structure of attractors that is much more homogeneous in the RC case than in the RS case
(compare Fig.~\ref{FIG2}(c,d) and (e,f)).
Moreover, the global angle of attractors is almost always along the `easy axes' of the $L\times L$ domain, 
and their number is 4 for large enough $L$ in the RC case (Fig.~\ref{FIG2}(a,g)). 
With RS disorder, attractors have more varied angles, and most often 2 are found in the accessible $L$ range, 
although their mean number slowly increases
with $L$ (Fig.~\ref{FIG2}(b,h)). 

\begin{figure}[t!]
	\includegraphics[width=\columnwidth]{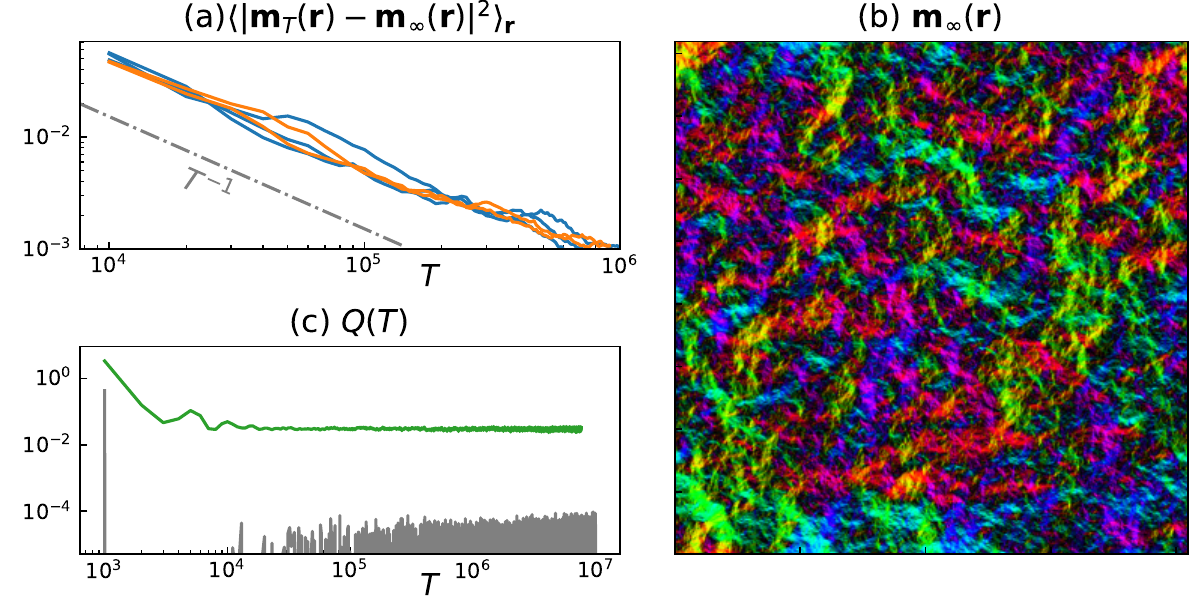}
	\caption{Ergodicity and memory in the disordered phase (RS case, $\eta=0.18$, $\varepsilon=0.055$, L=2048).
	(a) $\langle | {\bf m}_T({\bf r})-{\bf m}_\infty({\bf r})|^2\rangle_{\bf r}$ v $T$ for different initial 
	conditions, either ordered (orange curves) or taken in the steady state (blue curves). All converge  like $1/T$ 
	to ${\bf m}_\infty({\bf r})$.
	(b) long-time average momentum field ${\bf m}_\infty({\bf r})$ used in (a) (colormap as in Fig.~\ref{FIG2}).
	(c) $Q(T)$ v $T$ in the steady state (grey curve: pure case $\varepsilon=0$)).
	 }
	\label{FIG3}
\end{figure}

{\it Memory.}
Quenched disorder induces permanent memory of the underlying frozen landscape in both the ordered and disordered phases. This is best seen from the existence of well-defined non-trivial long-time averaged fields such as ${\bf m}_\infty({\bf r})$ and the fact that Edwards-Anderson-like order parameters such as
$Q_{\rm EA}= \lim_{T\to\infty} Q(T)$ with $Q(T)=\langle {\bf m}({\bf r},t)\cdot{\bf m}({\bf r},t+T)\rangle_{{\bf r},t}$ take finite values. 
In the disordered phase, ergodicity holds and all initial conditions eventually lead to the same 
long-time dynamics and momentum field ${\bf m}_\infty({\bf r})$ (Fig.~\ref{FIG3}(a,b)).
In the steady state, $Q(T)$ converges to some small but finite, sample-dependent $Q_{\rm EA}$ value, 
in contrast with the disordered phase of the VM  ($\varepsilon=0$) for which $Q(T)$ fluctuates around zero (Fig.~\ref{FIG3}(c)). 
In the ergodicity-broken ordered phase, $Q_{\rm EA}$ 
takes rather large values that are not only sample-dependent, 
but also attractor-dependent in the RS case (not shown).  

%%%%%%%%%%%%%%%%%%%%%%%%%%%
%%%%%%%%%%%%%%%%%%%%%%%%%%%

The ergodicity and memory properties presented above were found in both types of 
quenched disorder considered.
However, as indicated by the layout of their phase diagrams and the structure and statistics of attractors,
the RC and RS cases are fundamentally different, as we now show more quantitatively. 

\begin{figure}[t!]
	\includegraphics[width=1.\columnwidth]{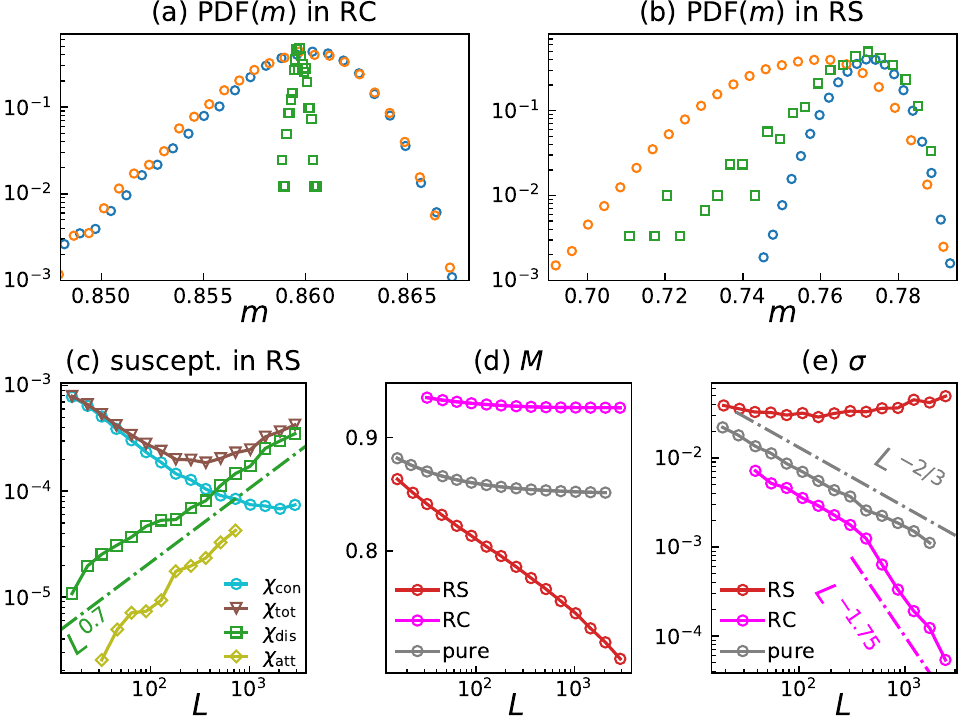}
	\caption{Fluctuations and order in the ordered phases (same parameters as in Fig.~\ref{FIG2}).
	(a): PDF($m$) for 2 attractors of the same sample (blue and orange) and 
	PDF($\langle m\rangle_t$) over samples and attractors (green) in the RC case ($L=512$).
	(b) same as (a) but for RS case ($L=512$)
	(c) $\chi_{\rm con}$, $\chi_{\rm dis}$, $\chi_{\rm att}$, and $\chi_{\rm tot}$ v $L$ for RS case. 
	(d) $M$ v $L$ for RC, RS, and VM ($\varepsilon=0$).
	(e) local exponent $\sigma$ v $L$ (extracted from data in (d), calculated as 
	$\sigma(\sqrt{L_nL_{n+1}})=-\log(M(L_{n+1})/M(L_n))/\log(L_{n+1}/L_n)$ 
	where $L_{n,n+1}$ are 2 consecutive system sizes).
	 }
	\label{FIG4}
\end{figure}

{\it Fluctuations in the ordered phase.} 
The breakdown of ergodicity in polarly ordered phases implies to consider 3 sources of fluctuations: 
dynamical, sample-to-sample, and attractor-to-attractor, illustrated in Fig.~\ref{FIG4}.
For a given attractor of a given sample, $m$ fluctuates in time, 
yielding an asymmetric probability distribution function (${\rm PDF}(m)$). 
Attractors of a given sample give near-identical 
${\rm PDF}(m)$ in the RC case, but not in the RS case 
(Fig.~\ref{FIG4}(a,b)). 
The sample- and attractor-averaged ${\rm PDF}(\langle m\rangle_t)$ is a very narrow Gaussian in the RC case, 
but is wide and asymmetric in the RS case
(green symbols in Fig.~\ref{FIG4}(a,b)). This means that, for the RC case presented, not only attractor-to-attractor 
but also sample-to-sample fluctuations are negligible compared to dynamical ones. 
In the RS case of Fig.~\ref{FIG4}(b), on the other hand, neither source of fluctuations 
can be neglected a priori.

To gauge which fluctuations will dominate and the nature of orientational order in the $L\to\infty$ limit, 
we now turn to finite-size effects.
We define the following
`connected' (dynamical), `disconnected' (sample-to-sample), and `attractor' susceptibilities
\footnote{For convenience, we omit the usual $L^2$ factor here.}:
\begin{subequations}
\begin{align}
\label{sus-con}
\chi_{\rm con} &= \overline{\left[ \langle m^2 \rangle - \langle m \rangle^2 \right]} \\
\label{sus-dis}
\chi_{\rm dis} &=  [\overline{\langle m\rangle}^2]  - [\overline{\langle m \rangle}]^2 \\
\label{sus-rep}
\chi_{\rm att} &= [ \overline{\langle m\rangle^2} - {\overline{\langle m \rangle}^2} ] 
\end{align}
\label{eq:sus}
\end{subequations}
where angle and square brackets respectively stand for averages over time and samples, 
while the upper bar denotes average over attractors.

In the RC case, 
$\chi_{\rm con}$ remains dominant at all sizes, and decreases fast with increasing $L$ (not shown):
this phase is self-averaging. 
Strikingly, in the RS case,
both $\chi_{\rm dis}$ and $\chi_{\rm att}$ grow like $L^\alpha$ with $\alpha\sim 0.7$,
while $\chi_{\rm con}$ decreases and seems to level off (Fig.~\ref{FIG4}(c)). 
This divergence of $\chi_{\rm dis}$ and $\chi_{\rm att}$ means that the system is strongly non-self-averaging 
\footnote{A self-averaging system would have $\alpha=-2$; weak self-averaging ($\alpha<0$) is also possible
see, e.g., \cite{vojta2006rare}.}
and that sample-to-sample 
fluctuations will dominate asymptotically (since $\chi_{\rm att}\ll \chi_{\rm dis}$, assuming this behavior holds for $L\to\infty$). 
As a result, for the system presented in Fig.~\ref{FIG4}(c), the total susceptibility
$\chi_{\rm tot}\equiv\chi_{\rm con}+\chi_{\rm dis}+\chi_{\rm att}\simeq \chi_{\rm con}+\chi_{\rm dis}$ 
first decreases with $L$ but then increases at large sizes when $\chi_{\rm dis}$ dominates.

%%%%%%%%%%%%%%%%%%%%%%%%%%%
%%%%%%%%%%%%%%%%%%%%%%%%%%%

Strong non-self-averaging 
implies that estimating numerically the scaling of the main global polar order parameter
$M=\overline{[\langle m\rangle]}$ is numerically challenging in the RS case. 
Fig.~\ref{FIG4}(d) shows $M(L)$ for the same parameters as in the rest of the figure, after averaging over typically 1000 samples and recording $m$ for millions of timesteps after transients 
(see numerical details in \cite{Note3}). Whereas $M(L)$ decreases slower than a powerlaw
in the RC case, indicating true long-range polar order, it decreases faster in the RS case.
The local slope (exponent) $\sigma(L)$ of these loglog plots offers more insight (Fig.~\ref{FIG4}(e)). 
In the RC case, $\sigma$ first goes to zero as $L^{-\omega}$ with $\omega\simeq\tfrac{2}{3}$ 
as in the VM~\cite{chate2020dry,mahault2019TT},
but then adopts a steeper decay with $\omega\simeq 1.75$ beyond some crossover scale,
which we believe to be related to the system size at which non-steady samples vanish (cf. Fig.~\ref{FIG2}(g)).
That $\omega\neq\tfrac{2}{3}$ indicates that the RC long-range ordered phase is {\it not} a Toner-Tu liquid \cite{Note3}.
In the RS case, $\sigma$ first decreases slightly, levels off, but then increases: 
a simple quasi-long-range order (algebraic decay of $M$, constant $\sigma$) is excluded.

\begin{figure}[t!]
	\includegraphics[width=\columnwidth]{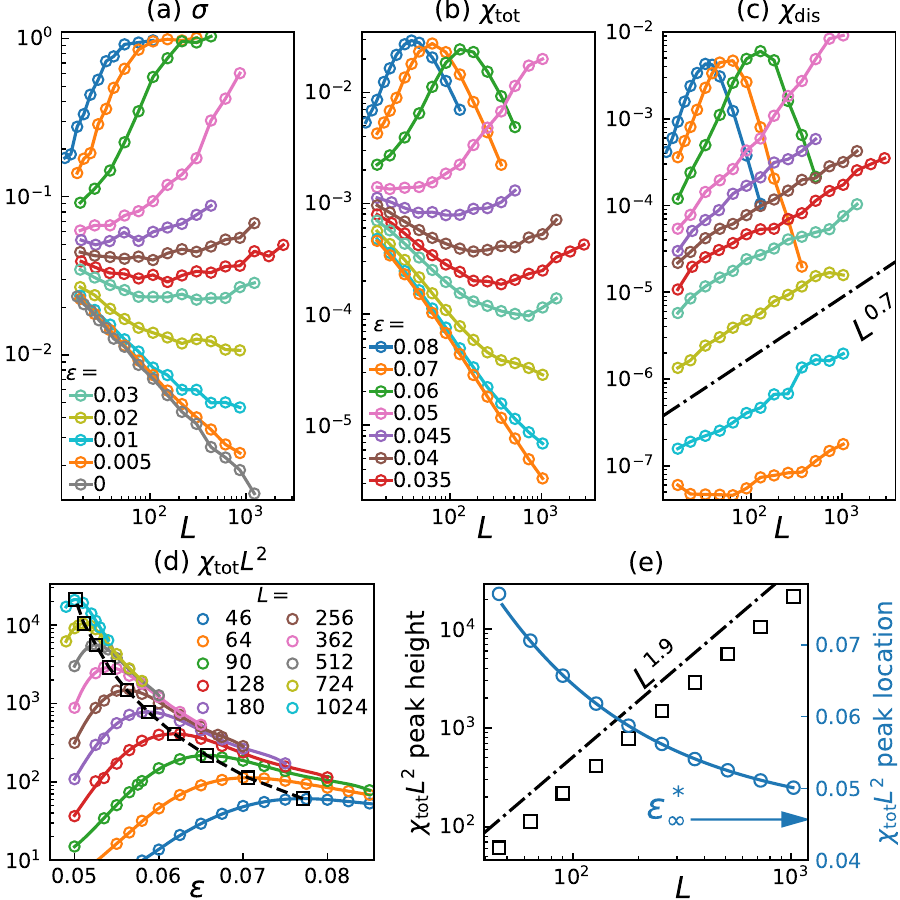}
	\caption{RS case at $\rho_0=1$, $\eta=0.18$.
	(a-c) local slope $\sigma$, $\chi_{\rm tot}$, and $\chi_{\rm dis}$ v $L$ for various $\varepsilon$ values 
	(see legends in (a,b)). 
	(d) $\chi_{\rm tot}\times L^2$ v $\varepsilon$ across the order/disorder transition at various sizes. 
	The estimated peaks are indicated by the black open squares.
	(e) scaling of the peaks detected in (d): heights (black squares, left scale) 
	and locations $\varepsilon^*(L)$ (blue circles, right scale). The solid blue line is a fit
	$\varepsilon^*(L)-\varepsilon^*_\infty\sim L^{-1/\nu}$ with $\nu\simeq 1.66$ and 
	$\varepsilon^*_\infty\simeq 0.0453$. 
	 }
	\label{FIG5}
\end{figure}

{\it Asymptotic nature of the quasi-ordered phase in the RS case.}
Scanning the whole phase at fixed $\rho_0$ and $\eta$ varying $\varepsilon$ clarifies the situation
described above at a single $\varepsilon$ value without bringing definitive answers. 
Fig.~\ref{FIG5}(a,b) show $\sigma(L)$ and $\chi_{\rm tot}(L)$ at various $\varepsilon$ values. 
At very small $\varepsilon$, $\sigma$ and $\chi_{\rm tot}$ first decay with $L$ like in the VM, 
then depart from this trend at a crossover scale which decreases with increasing $\varepsilon$. 
In line with the data in Fig.~\ref{FIG4} obtained for a particular $\varepsilon$ value, 
we find no evidence at any $\varepsilon$
of `simple' quasi-long-range order in the range of scales studied: Once $\sigma$ has stopped decreasing, 
it does not really plateau and starts increasing slowly. 

%%%%%%%%%%%%%%%%%%%%%%%%%%%
%%%%%%%%%%%%%%%%%%%%%%%%%%%

At the largest $\varepsilon$ values considered, $\sigma$ and $\chi_{\rm tot}$ increase with $L$, 
then $\sigma$ levels off at $+1$, the value
characteristic of the short-range order of the disordered phase ($M\sim1/L$), while  $\chi_{\rm tot}$ decreases. 
The maximum of $\chi_{\rm tot}$ is a measure of the correlation length, which seems to diverge 
when $\varepsilon$ decreases. 
Correspondingly,  $\chi_{\rm tot}L^2$ exhibits a maximum as a function of $\varepsilon$, 
whose height scales as $L^{\gamma/\nu}$ with $\frac{\gamma}{\nu}\simeq 1.9$, while its location 
$\varepsilon^* (L)$ seems to converge to a finite asymptotic value (Fig.~\ref{FIG5}(d,e)).
All this points to a continuous phase transition separating the disordered phase from the quasi-ordered one.
But 
sample-to-sample fluctuations diverge with $L$ all over the quasi-ordered regimes
with the same exponent as reported above, $\chi_{\rm dis}\sim L^{0.7}$ (Fig.~\ref{FIG5}(c)). 
These fluctuations having barely started to dominate $\chi_{\rm tot}$ at the largest accessible scales (Fig.~\ref{FIG5}(b)), 
it is premature to conclude about the asymptotic nature of the phase corresponding to the
quasi-ordered states observed at finite size, 
and a fortiori about the transition.

To summarize, 
quenched disorder affects polar active matter in more ways 
than believed so far. 
In particular, it breaks ergodicity in the ordered regimes observed at finite size, but not in the
disordered phase, which only shows infinite memory of the frozen landscape. 
We also showed that the ordered phase depends qualitatively 
on the type of quenched disorder: for random couplings,
it remains long-range ordered, but differently from the pure case. 
For random scatterers, we find strong non-self-averaging, with
sample-to-sample fluctuations dominating asymptotically. 
Unfortunately, the nature of this asymptotic regime remains largely inaccessible numerically.

We have started exploring other implementations of quenched disorder, such as random field, dilute scatterers, and
we find that their properties similar to the RS case. Moreover,
it is relatively easy to build systems similar to the RC case presented here \cite{TBP}.
We thus believe the two cases studied represent large classes of disordered active systems. 

Some of our results should be observable experimentally, e.g. in the Quincke roller system of 
\cite{bricard2013emergence,bricard2015emergent,geyer2018sounds}, which is believed
to be a realization of (effectively) dry polar active matter. However, our findings seem to contradict the conclusions
of \cite{chardac2020meandering}: there the breakdown of polar flocks was argued to lead to a ``dynamical vortex glass'' with many coexisting attractors, while we have shown that sufficiently long averages reveal an ergodic disordered phase
with infinite memory (Fig.~\ref{FIG3}). 
We hope that  further experiments will clarify this important point.

\acknowledgments
We thank Denis Bartolo, Gilles Tarjus and Matthieu Tissier for interesting discussions, and Julien Tailleur
for a critical reading of a previous version of this work. 
We acknowledge generous allocations of cpu time on the Living Matter Department cluster in MPIDS, 
on Beijing CSRC’s Tianhe supercomputer, 
and from the High-Performance Computing Center of Collaborative Innovation Center of Advanced Microstructures in Nanjing.
This work is supported by the National Natural Science Foundation of China 
(Grants \# 11635002 to X.-q.S. and H.C., 11922506 and 11674236 to X.-q.S.,
11974175 and 11774147 to Y.-q.M.).

\bibliography{Biblio-quenched.bib}

\end{document}